\documentclass[letterpaper,english,reprint, aps]{revtex4-1}
\usepackage[T1]{fontenc}
\usepackage[latin9]{inputenc}
\usepackage{units}
\usepackage{amsmath}
\usepackage{amssymb}
\usepackage{graphicx}



\usepackage{babel}
\begin{document}
\title{Fluid infiltration of a heterogeneous medium: A stochastic model}
\author{Clinton DeW. Van Siclen}
\email{cvansiclen@gmail.com}

\address{1435 W 8750 N, Tetonia, Idaho 83452, USA}
\date{26 March 2022}
\begin{abstract}
Fluid infiltration of a permeable brick in contact with a pressurized
reservoir of fluid is considered. A stochastic model, informed by
Darcy's law and the incompressibility of the fluid, shows how the
heterogeneity of the permeability field affects the time evolution
of the fluid infiltration. In particular, the cause of ``anomalous''
(non-Darcian) advance of a plume is determined. The model is applied
to bricks that are linear arrays of Sierpinski carpets. These calculated
results are compared to experimental results available in the literature,
to verify the model and method.
\end{abstract}
\maketitle

\section{Introduction}

Fluid flow through heterogeneous media is a topic of continuing scientific
interest with important technological applications and implications.
The particular phenomenon considered here is fluid infiltration of
a permeable brick due to contact, at one end of the brick, with a
pressurized reservoir of the fluid. Not only the heterogeneity of
the brick composition but also the confining surfaces of the brick
affect the rate of advance of the fluid into the brick.

The operative equation for this is Darcy's law, which relates the
pressure gradient in the fluid to the flow rate. The \textit{local}
version is
\begin{equation}
\mathbf{q}(\mathbf{r})=-\frac{\kappa(\mathbf{r})}{\mu}\mathbf{\mathbf{\nabla}}p(\mathbf{r})\label{eq:1}
\end{equation}
where $\mathbf{q}(\mathbf{r})$ is the volumetric flow rate at position
$\mathbf{r}$ within the brick, $p(\mathbf{r})$ is the fluid pressure,
$\kappa(\mathbf{r})$ is the permeability, and $\mu$ is the viscosity
of the fluid. {[}Note that other driving forces, such as gravity and
capillarity, are not considered here.{]} The incompressibility of
the fluid produces the additional equation
\begin{equation}
\mathbf{\mathbf{\nabla}}\cdot\mathbf{q}(\mathbf{r})=0\label{eq:2}
\end{equation}
which equivalently (for the purposes of this paper) means there are
no fluid sources or sinks except at the boundaries of the fluid plume.

\begin{figure}[b]
\includegraphics[scale=0.8]{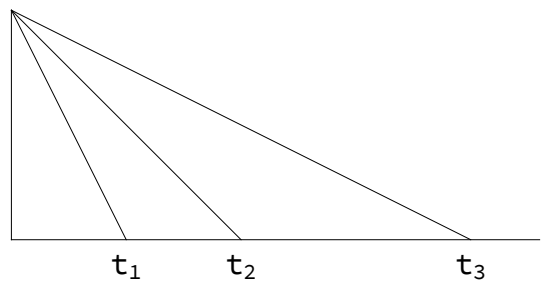}

\caption{Fluid pressure profiles, at successive times $t$, within a plume
infiltrating a one-dimensional, homogeneous brick. The fluid pressure
at the reservoir/brick interface (represented by the vertical line)
remains constant as the plume advances left to right, while the pressure
gradient within the plume remains linear. In effect, the ``steady-state''
condition holds \textit{continuously} as the plume grows.\label{Fig1}}
\end{figure}

A realistic model for fluid advance in a one-dimensional (1D) homogeneous
brick is illustrated in Fig. \ref{Fig1}.
 It shows the fluid pressure profile in the brick, at three successive
times $t_{1}<t_{2}<t_{3}$. As the fluid is incompressible, the steady-state
condition Eq. (\ref{eq:2}) holds continuously so that the pressure
profile remains linear, with slope (gradient) $-p(0)/x(t)$ where
$p(0)$ is the constant fluid pressure at the reservoir/brick interface,
and $x(t)$ is the extent of the fluid advance into the brick at time
$t$. The flow rate into the brick at time $t$ is $d\left[x(t)\right]/dt$.
Then Darcy's law implies
\begin{equation}
\frac{d}{dt}x(t)=\frac{\kappa}{\mu}\:\frac{p(0)}{x(t)}\label{eq:3}
\end{equation}
which has the solution

\begin{equation}
x(t)=\left[2\,\frac{\kappa}{\mu}\:p(0)\right]^{1/2}t^{1/2}.\label{eq:4}
\end{equation}
It bears comment that the value of the reservoir pressure $p(0)$
has no effect on the time exponent.

This model is the motivation for what follows. Most importantly, it
relies on the continuous application of the steady-state condition
as the fluid advances.

To apply the model to 2D and 3D heterogeneous bricks, then, it is
necessary to ensure a steady-state pressure field $\left\{ p(\mathbf{r})\right\} $
in the infiltrated fluid at time $t$. This is done by the Walker
Diffusion Method (WDM), which for this purpose effectively solves
the large set of coupled equations that are the discretized version
of the steady-state equation
\begin{equation}
\nabla\cdot\left[\frac{\kappa(\mathbf{r})}{\mu}\mathbf{\nabla}p(\mathbf{r})\right]=0\label{eq:5}
\end{equation}
with boundary conditions of constant pressure at the reservoir/brick
interface, and zero pressure ahead of the fluid front. Such calculations
performed for a succession of times $t$ gives the time evolution
of the fluid infiltration (the size of the plume).

Of course the pressure field and fluid advance are influenced by the
domain (referring to regions of homogeneous permeability) geometry
and composition of the brick. It is generally believed that the time
evolution is a power law, with an exponent $\gamma$ that reflects
the physical character of the portion of the brick saturated by the
fluid. In cases that a brick contains regions of different permeability,
the exponent value may be calculated by the WDM (via the pressure
field calculations), as demonstrated in this paper. Indeed, the primary
goal for this model is to connect $\gamma$ values with brick domain
geometries, so that experimentally obtained values can be interpreted.

The WDM is described briefly in the following section; subsequent
sections develop the fluid infiltration model and apply it to illustrative
bricks. A final section very briefly summarizes the general results
that are obtained.

\section{Walker Diffusion Method}

The WDM {[}\citealp{PRE99},\citealp{PRE02}{]} exploits the isomorphism
between the transport equations {[}e.g., Eq. (\ref{eq:1}){]} and
the diffusion equation for a collection of noninteracting random walkers
in the presence of a driving force,
\begin{equation}
\mathbf{J}(\mathbf{r})=-D(\mathbf{r})\,\rho(\mathbf{r})\mathbf{\,\nabla}\phi(\mathbf{r})\label{eq:6}
\end{equation}
where $D(\mathbf{r})$ and $\rho(\mathbf{r})$ are the local walker
diffusion coefficient and local walker density, respectively. In this
application of the WDM, the product $D(\mathbf{r})\,\rho(\mathbf{r})$
is identified with the local permeability $\kappa(\mathbf{r})$. Specifically,
$D(\mathbf{r})\,\rho^{0}(\mathbf{r})=\kappa(\mathbf{r})/\mu$, where
$\rho^{0}(\mathbf{r})$ is the \textit{equilibrium} walker density
(meaning, in the absence of a driving force). So that the walker trajectories
fully reflect the domain geometry the local diffusion coefficient
$D(\mathbf{r})$ is everywhere set equal to $1$.

The domains {[}distinguished by different values of $\kappa(\mathbf{r})/\mu${]}
of the medium thus correspond to distinct populations of walkers,
where the equilibrium walker density of a population is given by the
value $\kappa(\mathbf{r})/\mu$. The principle of detailed balance
ensures that the equilibrium population densities are maintained,
and so provides the following rule for walker diffusion over a digitized
(pixelated) multi-domain material: a walker at site (or pixel) $i$
attempts a move to a randomly chosen adjacent site $j$ during the
time interval $\tau=(4d)^{-1}$, where $d$ is the Euclidean dimension
of the space; this move is successful with probability $p_{i\rightarrow j}=\kappa_{j}/(\kappa_{i}+\kappa_{j})$,
where $\kappa_{i}$ and $\kappa_{j}$ are the permeabilities of sites
$i$ and $j$, respectively. (Note this is ``blind ant'' behavior.)
The path of a walker thus reflects the permeability and morphology
of the domains that are encountered.

For the purposes of this paper, a key component of the WDM {[}\citealp{PRE02}{]}
is the concept of a ``residence time'' associated with each site
visited by a walker. As the walker diffuses over the permeable medium,
the time required for a move from site $i$ (which may be many multiples
of $\tau$) is accrued to the residence time $t_{i}$. Note that at
equilibrium (i.e., in the absence of a driving force), a single diffusing
walker will occupy the sites of the system in proportion to the corresponding
transport coefficients $\kappa(\mathbf{r})/\mu$; thus, in the limit
of infinite time, the \textit{equilibrium} walker density at site
$i$ is given by
\begin{equation}
\rho_{i}^{0}=\frac{\kappa_{i}}{\mu}=\frac{t_{i}}{\left\langle t_{k}\right\rangle }\frac{\left\langle \kappa_{k}\right\rangle }{\mu}\label{eq:7}
\end{equation}
where the averages implied by the angle brackets are taken over all
visited sites $k$.

These walker densities are altered when a potential gradient is created
by injecting numerous walkers into the system at one boundary or point
(the ``source''), and removing them at another (the ``sink'').
The \textit{steady-state} walker density at site $i$ is then
\begin{equation}
\rho_{i}=\frac{t_{i}}{\left\langle t_{k}\right\rangle }\frac{\left\langle \kappa_{k}\right\rangle }{\mu}\label{eq:8}
\end{equation}
where the residence time $t_{i}$ accounts for visits by \textit{all}
walkers to site $i$, and the averages implied by the angle brackets
are taken over all visited sites $k$. As derived in Ref. {[}\citealp{PRE02}{]},
the (dimensionless) potential at site $i$ is
\begin{equation}
\phi_{i}=\frac{\rho_{i}}{\rho_{i}^{0}}=\frac{t_{i}}{\left\langle t_{k}\right\rangle }\frac{\left\langle \kappa_{k}\right\rangle }{\kappa_{i}}\label{eq:9}
\end{equation}
and the walker flux $J_{i\rightarrow j}$ between adjacent sites is
\begin{equation}
J_{i\rightarrow j}=\frac{1}{\mu}\:\frac{2\,\kappa_{i}\,\kappa_{j}}{\kappa_{i}+\kappa_{j}}\left(\phi_{i}-\phi_{j}\right).\label{eq:10}
\end{equation}

In this paper, the driving force for fluid flow is the fluid pressure
gradient, so the potential field $\left\{ \phi_{i}\right\} $ corresponds
to the fluid pressure within the plume. Like the path of a single
walker, the potential field $\left\{ \phi_{i}\right\} $ {[}and so
the pressure field $\left\{ p_{i}\right\} ${]} reflects the composition
and morphology of the medium. Examples of electrical potential fields
calculated in this way are given in Ref. {[}\citealp{PRE02}{]}; there
they are shown to produce the correct values for the effective (macroscopic)
conductivity of multiphase composites.

Potential-field contour plots show the flow direction (perpendicular
to the contours) within the brick. Due to fluid incompressibility,
regions of lower (higher) permeability are distinguished by their
narrower (wider) contour spacing, indicating a larger (smaller) fluid
pressure gradient.

In the WDM calculations below, it is computationally advantageous
to utilize the variable residence time algorithm {[}\citealp{PRE99}{]},
which takes advantage of the statistical nature of the diffusion process.
According to this algorithm, the actual behavior of a walker is well
approximated by a sequence of moves in which the direction of the
move from a site $i$ is determined randomly by the set of probabilities
$\left\{ P_{i\rightarrow j}\right\} $, where $P_{i\rightarrow j}$
is the probability that the move is to adjacent site $j$ (which has
permeability $\kappa_{j}$) and is given by the equation
\begin{equation}
P_{i\rightarrow j}=\frac{\kappa_{j}}{\kappa_{i}+\kappa_{j}}\left[\sum_{k=1}^{2d}\left(\frac{\kappa_{k}}{\kappa_{i}+\kappa_{k}}\right)\right]^{-1}.\label{eq:11}
\end{equation}
The sum is over all sites adjacent to site $i$. The time interval
over which the move occurs is
\begin{equation}
T_{i}=\left[2\sum_{k=1}^{2d}\left(\frac{\kappa_{k}}{\kappa_{i}+\kappa_{k}}\right)\right]^{-1}.\label{eq:12}
\end{equation}
Note that this version of the variable residence time algorithm is
intended for orthogonal systems (meaning a site in a 3D system has
six neighbors, for example).

Interestingly, Eqs. (\ref{eq:9}) and (\ref{eq:10}) reveal that the
potential field $\left\{ \phi\right\} $ is determined by the permeability
\textit{ratios} $\left\{ \kappa_{\beta}/\kappa_{\alpha}\right\} $
of the domains, while the flux $J$ is determined by the particular
values of $\left\{ \kappa/\mu\right\} $ as well.

It must be emphasized that the WDM is a mathematical method\textemdash \textit{not}
a particle model of a physical process. The eponymous walker diffuses
over a pixelated medium according to particular rules, thereby ``solving''
the system of local transport equations associated with the set of
pixels. Specifically, the potential field $\left\{ \phi_{i}\right\} $
is obtained.

\section{Construction of the brick and plume}

A one-dimensional, WDM-compliant model of the reservoir-brick system
is the 1D array
\[
\frac{\kappa_{0}}{\mu}\;\frac{\kappa_{1}}{\mu}\;\frac{\kappa_{2}}{\mu}\;\frac{\kappa_{3}}{\mu}\ldots
\]
where site $j=0$, with permeability $\kappa_{0}=\infty$, is the
fluid reservoir, and sites $j>0$ constitute the brick. Site $j=1$,
with permeability $\kappa_{1}$, is the walker ``source'' (meaning,
where new walkers are placed). Note that the effect of $\kappa_{0}=\infty$
is that a walker entering site $j=0$ will not return to the brick.

A plume of size $S$ is bounded by the ``sink'' at site $j=S+1$.
Thus a walker that enters site $j=S+1$ is \textit{immediately} removed
(meaning, no residence time is accrued to that site).

The potential field $\left\{ \phi_{j}\right\} $ is calculated from
the residence times $t_{j}$ accrued as the walkers diffuse over the
sites $j\geq1$ prior to their disappearance into the sink or the
reservoir.

Two- and three-dimensional bricks are constructed similarly, except
that the brick surfaces parallel to the direction of fluid flow are
clad with non-permeable ($\kappa=0$) sites. The one addition to the
walker-diffusion procedure is that the reservoir/brick interface is
kept at uniform potential by placing the next walker at the interface
(source) site $i$ with smallest value $t_{i}/\kappa_{i}$.

Fluid infiltration in heterogeneous bricks is affected by the spatially
varying permeability of the brick. To account for this, a plume of
size $S$ is constructed in the following way: Consider that a plume
is a set of contiguous \textit{visited} sites. A new walker released
from the source diffuses until it is either absorbed at the reservoir/brick
interface, or enters an \textit{unvisited} site and so converts it
to a \textit{visited} site (and is then removed). This one-site-at-a-time
growth procedure is repeated until the plume attains size $S$. Due
to the walker diffusion rules Eqs. (\ref{eq:11}) and (\ref{eq:12})
imposed by the WDM, this plume reflects (conforms to) the heterogeneity
of the brick.

\section{Plume infiltration model}

This model relies on the potential field $\left\{ \phi\right\} $
calculated by the WDM. That provides the value $\left\langle \phi_{1}\right\rangle $,
and the value $\left\langle J_{1\rightarrow2}\right\rangle $ as calculated
by Eq. (\ref{eq:10}). These averages are taken over all ``source''
sites at the reservoir/brick interface. Here the subscript ``$1$''
indicates a source site, and the subscript ``$2$'' indicates the
adjacent, non-source site. The width of a 2D brick is $W$ sites.
The model of course applies to 1D and 3D bricks as well.

Consider the \textit{steady-state} system that is a plume of size
$S$. Then the fluid flux $J=S/t_{S}$ where $t_{S}$ is the time
interval needed for a complete fluid ``refresh'' of the saturated
volume (set of sites) $S$. Thus
\begin{equation}
J=\left\langle J_{1\rightarrow2}\right\rangle W\label{eq:13}
\end{equation}
and
\begin{equation}
t_{S}=\frac{S}{J}=\frac{S}{\left\langle J_{1\rightarrow2}\right\rangle W}.\label{eq:14}
\end{equation}
In addition, Darcy's law implies
\begin{equation}
\frac{J}{W}=\frac{\kappa_{\mathrm{eff}}}{\mu}\,\frac{\left\langle \phi_{1}\right\rangle }{\frac{S}{W}}\label{eq:15}
\end{equation}
where $\mathrm{\kappa}_{\mathrm{eff}}$ is the effective permeability
of the region between source and sink. Combining these equations gives
\begin{equation}
\frac{\kappa_{\mathrm{eff}}}{\mu}=\left(\frac{S}{W}\right)^{2}\frac{1}{\left\langle \phi_{1}\right\rangle t_{S}}=\frac{S}{W}\,\frac{\left\langle J_{1\rightarrow2}\right\rangle }{\left\langle \phi_{1}\right\rangle }\label{eq:16}
\end{equation}
 and the relation
\begin{equation}
S=W\left[\frac{\kappa_{\mathrm{eff}}}{\mu}\,\left\langle \phi_{1}\right\rangle t_{S}\right]^{1/2}.\label{eq:17}
\end{equation}
Comparison with the infiltration equation
\begin{equation}
S(t)=W\left[2\,\frac{\kappa_{\mathrm{eff}}}{\mu}\,p(0)\,t\right]^{1/2}\label{eq:18}
\end{equation}
shows $\left\langle \phi_{1}\right\rangle t_{S}=2\,p(0)\,t$, so the
infiltration time
\begin{equation}
t=\frac{1}{2\:p(0)}\left\langle \phi_{1}\right\rangle t_{S}.\label{eq:19}
\end{equation}
This relation between $t$ and $t_{S}$ ensures that a steady-state
potential (pressure) profile is maintained as the plume advances.

Further, Eq. (\ref{eq:19}) connects this stochastic model to the
actual (physical) infiltration time $t$.

Note that this derivation assumes that the effective permeability
$\kappa_{\mathrm{eff}}$ is unchanging as the plume advances. Thus
the ``anomalous'' growth behavior is observed when this condition
is not met. In that case the plume growth may be described by the
power-law relation,

\begin{equation}
S\propto W\left[\left\langle \phi_{1}\right\rangle t_{S}\right]^{\gamma}\label{eq:20}
\end{equation}
with exponent $\gamma\neq1/2$. An exponent $\gamma$ value greater
(less) than $1/2$ corresponds to increasing (decreasing) effective
permeability $\kappa_{\mathrm{eff}}$ as the plume advances. {[}It
is easy to see that this is the case. The value $\kappa_{\mathrm{eff}}$
associated with a plume of size $S$ is found by running a line of
slope $1/2$ through the plotted point $\left(\ln\left\langle \phi_{1}\right\rangle t_{S},\ln S\right)$.
Then the y-intercept of that line is $\ln W+\nicefrac{1}{2}\ln\left(\kappa_{\mathrm{eff}}/\mu\right)$.
Thus successive points for which the y-intercept is increasing (decreasing)
indicates the value $\mathrm{\kappa_{\mathrm{eff}}}$ is increasing
(decreasing) as the plume grows.{]}

A subsequent transition to ``normal'' growth behavior ($\gamma=1/2$)
occurs when further plume growth causes $\kappa_{\mathrm{eff}}$ to
acquire a reasonably constant value. It is convenient for later discussion
to introduce the length $\xi_{p}$, as the linear dimension of the
plume when the transition occurs.

Note that Eq. (\ref{eq:9}) shows $\left\langle \phi_{1}\right\rangle =2$
in the case of homogeneous bricks. More generally, the calculated
$\left\langle \phi_{1}\right\rangle $ value varies until the transition
to ``normal'' plume behavior is underway, whereupon $\left\langle \phi_{1}\right\rangle \approx2$.

\begin{figure}
\includegraphics[scale=0.8]{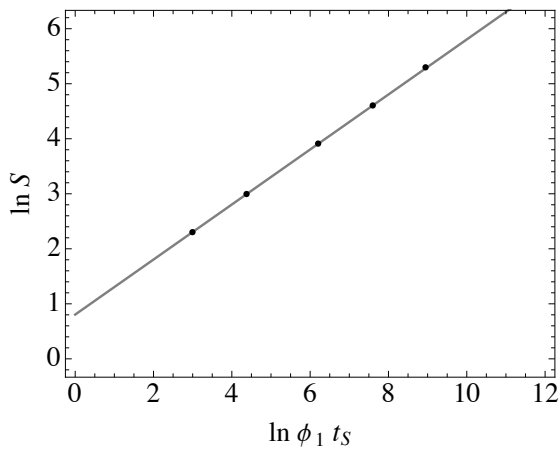}

\caption{Plume growth in a 1D homogeneous brick. The diagonal line has slope
$1/2$, indicating power-law growth with exponent $\gamma=1/2$.\label{Fig2}}
\end{figure}

\begin{figure}
\includegraphics[scale=0.8]{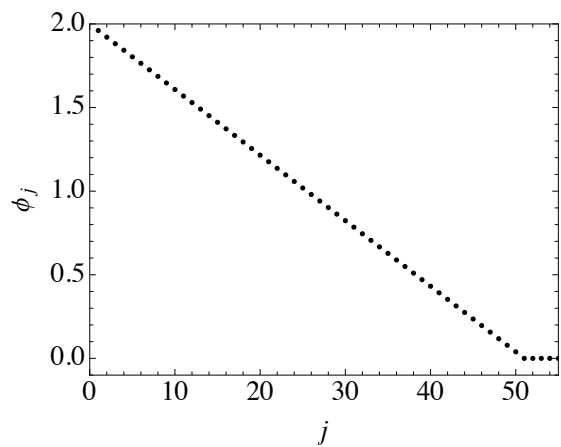}\caption{Potential profile $\left\{ \phi_{j}\right\} $ for the plume tracked
in Fig. \ref{Fig2}, at size $S=50$. The profile remains linear as
the plume grows.\label{Fig3}}
\end{figure}

As an example of the ``normal'' plume infiltration model, calculated
results for plume advance in a homogeneous (all $\kappa_{j}=5$) 1D
brick are presented in Figs. \ref{Fig2} and \ref{Fig3}. Figure \ref{Fig2}
is a log-log plot of $\phi_{1}\,t_{S}$ versus plume size $S$, showing
that exponent $\gamma=1/2$ in accord with Eq. (\ref{eq:17}). The
y-intercept $\ln W+\nicefrac{1}{2}\,\ln\left(\kappa_{\mathrm{eff}}/\mu\right)$
gives the effective permeability $\kappa_{\mathrm{eff}}=5$. Figure
\ref{Fig3} shows the potential profile $\left\{ \phi_{j}\right\} $
for a plume of size $S=50$. As expected, the gradient is constant
from source to sink.

In this example, and subsequent calculations, the fluid viscosity
$\mu=1$.

Note that the value $\mu$ is constant and uniform throughout a plume.
In addition, inspection of the WDM equations in Sec. II reveals that
all quantities are ultimately functions of ratios $\kappa/\mu$. Thus
fluid infiltration of a brick represented by the array of elements
$\left\{ \frac{\kappa_{j}}{\mu}\right\} $ resembles that for any
brick represented by an array having elements $\frac{\kappa'_{j}}{\mu'}=\frac{\kappa_{j}}{\mu}$.
For example, Fig. \ref{Fig2} is reproduced by a 1D brick represented
by an array of elements $\frac{\kappa_{j}}{\mu}=\frac{10}{2}$.

\section{Sierpinski carpets}

\begin{figure}
\includegraphics[scale=0.8]{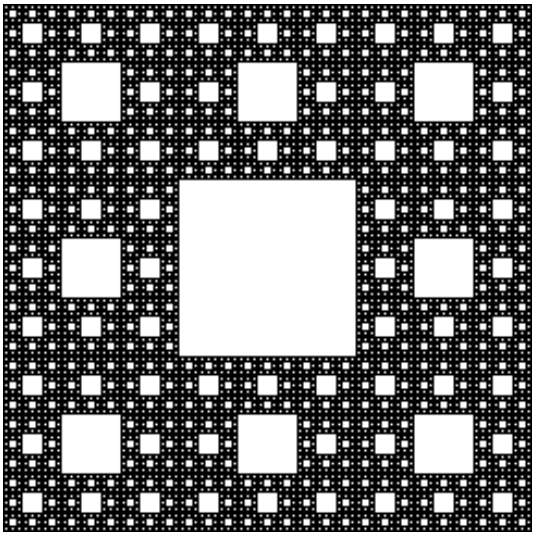}\caption{Fifth iteration Sierpinski carpet. The left edge of the carpet is
the reservoir/brick interface from which a plume grows. The connected
black domain has permeability $\kappa_{\alpha}$ while the distributed
white domains have permeability $\kappa_{\beta}$.\label{Fig4}}
\end{figure}

An interesting 2D brick is the Sierpinski carpet {[}\citealp{Sierp}{]},
which has a self-similar geometric structure comprising a connected
domain of permeability $\kappa_{\alpha}$ surrounding disconnected
domains of permeability $\kappa_{\beta}$. Figure \ref{Fig4} shows
the fifth iteration of the generator for the center-hole $(3,1)$
Sierpinski carpet, which is used in the calculations. There are five
sizes of white squares; the smallest white squares are the size of
a ``site''. Thus the carpet comprises $\left(3^{5}\right)^{2}=243^{2}$
sites. Each brick considered here is a linear array of carpets. The
fluid viscosity value $\mu=1$ is used in these calculations.

The potential field $\left\{ \phi\right\} $ is obtained by releasing
a very large number of walkers from the reservoir/brick interface
(the ``source''), and immediately removing them when they move over
the plume boundary (the ``sink''). For each brick considered below,
it is assumed that an accurate representation of the potential field
is achieved when $10^{6}$ walkers have reached the sink. Of course,
those walkers are accompanied by orders-of-magnitude more walkers
that are absorbed at the reservoir/brick interface.

The results for fluid infiltration are presented in log-log plots
of $\left\langle \phi_{1}\right\rangle t_{S}$ versus plume size $S$.
Points that don't lie precisely on a fitted line are believed to reflect
the details of the carpet geometry, not the stochastic nature of the
calculations. To aid comprehension, horizontal lines are added that
correspond to a saturated half-carpet (dotted line), and fully saturated
carpets (dashed lines) in the array.

\subsection*{Linear array of carpets with $\kappa_{\alpha}=1$ and $\kappa_{\beta}=0$}

\begin{figure}
\includegraphics[scale=0.8]{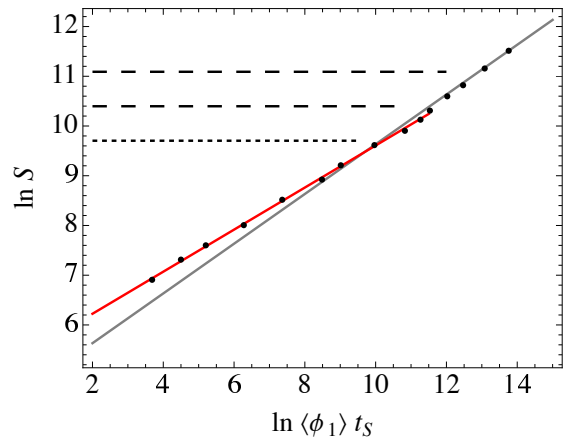}

\caption{Plume growth into a linear array of Sierpinski carpets with connected
domain $\kappa_{\alpha}=1$, and $\kappa_{\beta}=0$. The upper diagonal
line is fitted to points for plumes contained within the first carpet,
which reveals ``anomalous'' power-law growth with exponent $\gamma=0.423<1/2$.\label{Fig5}}

\end{figure}

Figure \ref{Fig5} is the $\log$-$\log$ plot of $\left\langle \phi_{1}\right\rangle t_{S}$
versus plume size $S$. The upper diagonal line is a fit to the points
for plumes that are contained within the first carpet. The slope of
that line is $\gamma=0.423<1/2$. The progressively smaller $\kappa_{\mathrm{eff}}$
values associated with the advancing (growing) plume can be calculated
by Eq. (\ref{eq:16}). The lower diagonal line has slope $1/2$ and
runs through the last point on the plot, obtained for a plume that
has entered the fourth carpet in the array. That last point gives
the effective permeability $\kappa_{\mathrm{eff}}=0.179$. The lower
horizontal dashed line at $S=3^{10}\left(8/9\right)^{5}$ corresponds
to a fully saturated single carpet; the upper horizontal dashed line
corresponds to a fully saturated two-carpet array.

\subsection*{Linear array of carpets with $\kappa_{\alpha}=1$ and $\kappa_{\beta}=0.01$}

\begin{figure}
\includegraphics[scale=0.8]{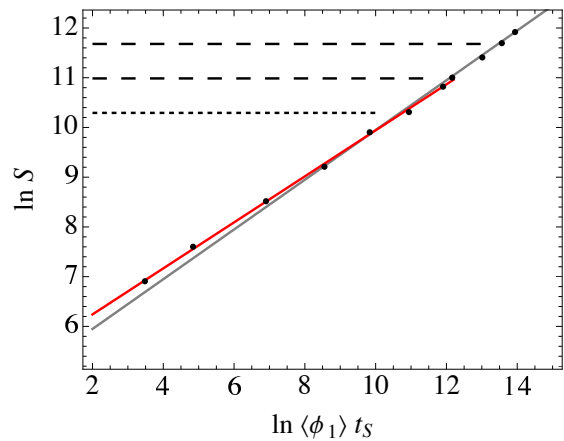}

\caption{Plume growth into a linear array of Sierpinski carpets with connected
domain $\kappa_{\alpha}=1$, and $\kappa_{\beta}=0.01$. The upper
diagonal line is fitted to points for plumes contained within the
first carpet, which reveals ``anomalous'' power-law growth with
exponent $\gamma=0.463<1/2$.\label{Fig6}}

\end{figure}

Figure \ref{Fig6} closely resembles Fig. \ref{Fig5}, the apparent
difference being the somewhat larger exponent value $\gamma=0.463$.
The lower diagonal line has slope $1/2$ and runs through the last
point on the plot, obtained for a plume that has entered the third
carpet in the array. That last point gives the effective permeability
$\kappa_{\mathrm{eff}}=0.336$.

\begin{figure}
\includegraphics[scale=0.8]{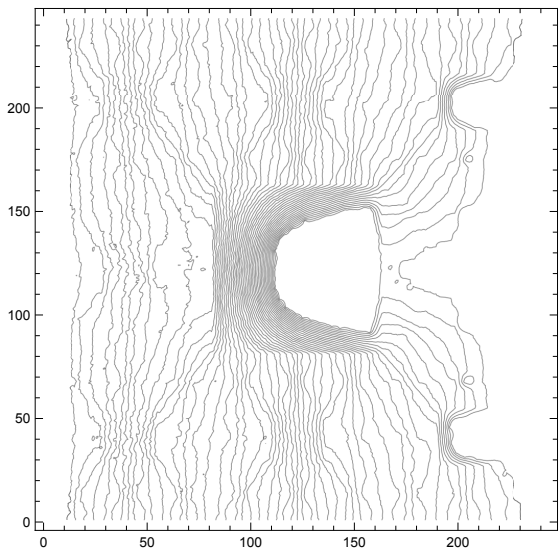}

\caption{Contour plot of the potential field $\left\{ \phi\right\} $ for a
plume of size $S=50000$ contained within a carpet with connected
domain $\kappa_{\alpha}=1$, and $\kappa_{\beta}=0.01$. This plume
contributed a point to Fig. \ref{Fig6}.\label{Fig7}}

\end{figure}

Figure \ref{Fig7} is a contour plot of the potential field for a
plume of size $S=50000$. The fluid reservoir, kept at constant pressure,
is at the left edge of the carpet. Note the steep potential gradient
produced in the low-permeability regions. Clearly those regions impede
and distort the plume's advance. However, once a low-permeability
domain is saturated, it appears to be less of an impediment to subsequent
fluid flow.

\subsection*{Linear array of carpets with $\kappa_{\alpha}=1$ and $\kappa_{\beta}=100$}

\begin{figure}
\includegraphics[scale=0.8]{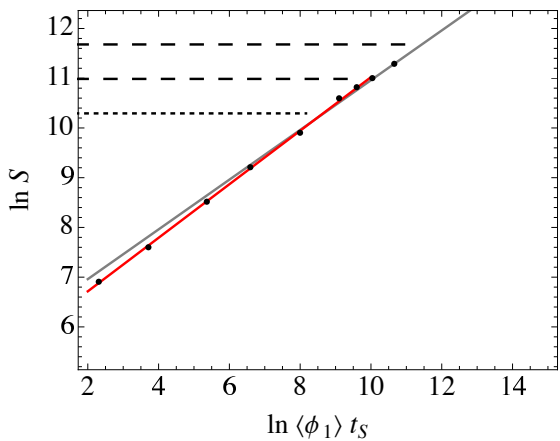}

\caption{Plume growth into a linear array of Sierpinski carpets with connected
domain $\kappa_{\alpha}=1$, and $\kappa_{\beta}=100$. The lower
diagonal line is fitted to points for plumes contained within the
first carpet, which reveals ``anomalous'' power-law growth with
exponent $\gamma=0.538>1/2$.\label{Fig8}}
\end{figure}

Figure \ref{Fig8} is the log-log plot of $\left\langle \phi_{1}\right\rangle t_{S}$
versus plume size $S$. The lower diagonal line is a fit to the points
for plumes that are contained within the first carpet. The slope of
that line is $\gamma=0.538>1/2$. The upper diagonal line has slope
$1/2$ and runs through the last point on the plot, obtained for a
plume that has entered the second carpet in the array. That last point
gives the effective permeability $\kappa_{\mathrm{eff}}=2.547$.

\begin{figure}
\includegraphics[scale=0.8]{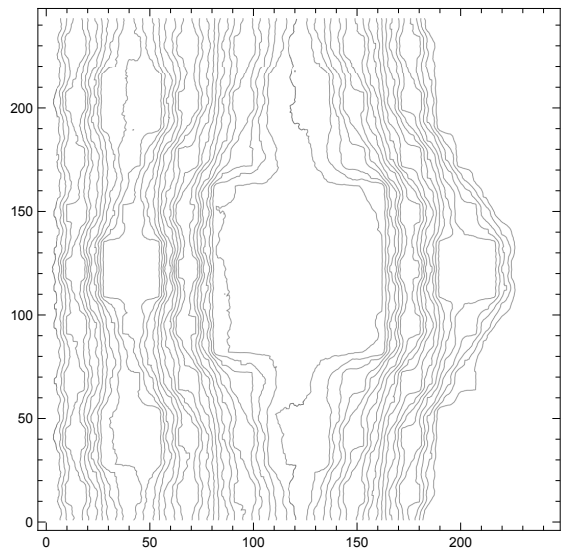}

\caption{Contour plot of the potential field $\left\{ \phi\right\} $ for a
plume of size $S=50000$ contained within a carpet with connected
domain $\kappa_{\alpha}=1$, and $\kappa_{\beta}=100$. This plume
contributed a point to Fig. \ref{Fig8}.\label{Fig9}}
\end{figure}

Figure \ref{Fig9} is a contour plot of the potential field for a
plume of size $S=50000$. The fluid reservoir, kept at constant pressure,
is at the left edge of the carpet. Note the very modest potential
gradient produced in the high-permeability regions.

Inspection of Figs. \ref{Fig5}, \ref{Fig6}, and \ref{Fig8} finds
that $\xi_{p}$ is not less than one carpet length (measured from
the reservoir/brick interface), in each case. In any event the transition
implied by $\xi_{p}$ may be rather gradual. That is, the transition
to ``normal'' growth behavior ($\gamma=1/2$) may not be complete
until the plume has saturated multiple carpets in the array. Indeed
this is true for the examples above: Denote by the function $\kappa\left(\kappa_{\alpha},\kappa_{b}\right)$
the effective permeability $\kappa_{\mathrm{eff}}$ of an infinite,
linear array of Sierpinski carpets comprised of domains with permeabilities
$\kappa_{\alpha}$ and $\kappa_{\beta}$. Then according to Ref. \citep{Chess},
\begin{equation}
\kappa\left(1,0.01\right)\:\kappa\left(1,100\right)=1.\label{eq:20a}
\end{equation}
Thus one or both of the calculated $\kappa_{\mathrm{eff}}$ values
reported above are too low (their product is $0.856$), so indicating
that ``normal'' growth behavior has not yet been achieved in one
or both systems. {[}Unfortunately, walkers get trapped in the non-percolating
$\kappa_{\beta}=100$ domains for long periods of computer time, making
calculations for arrays of many carpets infeasible. More points in
Fig. \ref{Fig8} would undoubtably reveal the true value $\kappa_{\mathrm{eff}}>2.547$.{]}

Although a precise numerical value for $\xi_{p}$ is impossible to
obtain, it is a useful concept.

\subsection*{Linear array of carpets with connected pore space ($\kappa_{\alpha}=\infty$)
and $\kappa_{\beta}=0$}

The walker behavior, and so the potential field $\left\{ \phi\right\} $,
are determined by the domain morphology (geometry) of the brick, and
by the \textit{ratios} of the different permeability values of the
domains comprising a brick. Thus application of the fluid infiltration
model to the Sierpinski carpet with connected pore space is identical
to that for the Sierpinski carpet with permeabilities $\kappa_{\alpha}=1$
and $\kappa_{\beta}=0$. Consequently the ``anomalous'' exponent
value $\gamma=0.423$ is the same in both cases.

This value is identical to that obtained experimentally by Filipovitch
\textit{et al}. {[}\citealp{Filip}{]} using a Hele-Shaw cell containing
a 3D-printed distribution of flow obstacles in the pattern of the
\textit{second} iteration of the generator for the center-hole (3,1)
Sierpinski carpet. (This is the pattern of Fig. \ref{Fig4}, but with
two, rather than five, sizes of white squares.) The fluid was pure
glycerin, under a fixed pressure head at one edge of the square carpet.
Further, Ref. {[}\citealp{Filip}{]} and Voller {[}\citealp{Voller}{]}
report simulations (thus an approach different from that used in this
paper) that obtain essentially this \textit{same} value of $\gamma$
for carpets in the pattern of the first, second, third, and fourth
iterations. Subsequently, variations on those simulations were made
by Aar$\tilde{\mathrm{a}}$o Reis \textit{et al}. {[}\citealp{Aarao}{]}
to better understand the carpet-plume system.

That the power-law exponent $\gamma$ is the same for carpet iterations
$i\leq5$ is due to the fact that as the plume front moves from the
reservoir/brick ``source'' through the $i=5$ carpet, it is passing
through the $i<5$ carpets in turn (see Fig. \ref{Fig4}). Thus this
is an effect of the self-similar character of the Sierpinski carpet,
on the fluid infiltration.

Note that a Sierpinski carpet with connected pore space can be regarded
as a homogeneous material with an effective permeability $\kappa_{\mathrm{eff}}$.
Clearly that value reflects the porosity $\left(8/9\right)^{i}$ of
the carpet. Thus $\kappa_{\mathrm{eff}}$ will decrease as the plume
advances through the $i=5$ carpet by filling the $i<5$ carpets in
turn. That sequence produces a time-evolution exponent $\gamma<1/2$.

In fact $\kappa_{i}$ will decrease \textit{faster} than $\left(8/9\right)^{i}$,
since the porosity value does not fully account for the distribution
of obstacles-to-flow within the carpet. Here $\kappa_{i}$ is the
effective permeability for a fully saturated, iteration $i$ Sierpinski
carpet. A crude upper bound for the $\gamma$ value is obtained as
follows.

Reference {[}\citealp{Sierp}{]} provides the relation
\begin{equation}
\frac{S_{i}}{W_{i}}=\left(\frac{8}{9}\right)^{i}W_{i}=\left[\left(\frac{8}{9}\right)^{i}\right]^{1-\nu}\label{eq:21}
\end{equation}
where the subscript $i$ refers to the iteration $i$ Sierpinski carpet,
and $\nu\equiv\left(2-\mathcal{H}\right)^{-1}$ where $\mathcal{H=\ln}8/\ln3$
is the Hausdorff (fractal) dimension of the Sierpinski carpet. Then
use of Eq. (\ref{eq:20}) in Eq. (\ref{eq:16}) gives
\begin{equation}
\frac{\kappa_{i}}{\mu}=\left(\frac{S_{i}}{W_{i}}\right)^{2}\frac{1}{\left\langle \phi_{1}\right\rangle t_{S}}\propto\left(\frac{S_{i}}{W_{i}}\right)^{2}\left(\frac{S_{i}}{W_{i}}\right)^{-1/\gamma}\label{eq:22}
\end{equation}
thereby producing the relation
\begin{equation}
\frac{\kappa_{i}}{\mu}\propto\left[\left(\frac{8}{9}\right)^{i}\right]^{\eta}\label{eq:23}
\end{equation}
with exponent $\eta$ being 
\begin{equation}
\eta\equiv\left(1-\nu\right)\left(2-\frac{1}{\gamma}\right).\label{eq:24}
\end{equation}
In order that $\kappa_{i}/\mu$ decrease faster than $\left(8/9\right)^{i}$,
the exponent $\eta$ must be greater than $1$. After some algebra,
the upper bound is obtained:
\begin{equation}
\gamma<\left(2+\frac{1}{\nu-1}\right)^{-1}=1-\frac{1}{\mathcal{H}}=0.471679\label{eq:25}
\end{equation}

\section{Concluding remarks}

The stochastic model and method developed here for fluid infiltration
of a permeable brick utilizes a \textit{local} version of Darcy's
law in order to account for the heterogeneities in the brick. For
plumes that have not yet advanced a distance $\xi_{p}$ into the brick,
the time evolution is ``anomalous'', caused by significant variation
in the effective permeability $\mathrm{\kappa}_{\mathrm{eff}}$ as
the plume grows. Specifically, an exponent $\gamma$ value greater
(less) than $1/2$ is produced by growth into regions of generally
increasing (decreasing) permeability. Plumes that have advanced beyond
$\xi_{p}$ show ``normal'' behavior, indicated by the exponent value
$\gamma=1/2$ (which in fact is a consequence of having little or
no variation of $\kappa_{\mathrm{eff}}$ as the plume grows).

Here the method is applied to bricks that are linear arrays of Sierpinski
carpets. These are fairly representative bricks, in that they comprise
a connected (percolating) domain of permeability $\kappa_{\alpha}$
surrounding isolated domains of permeability $\kappa_{\beta}$. Plume
growth within the first carpet presents as a power law with exponent
$\gamma\neq1/2$, in accordance with the plume infiltration model
in Sec. IV.

The contour plots of the potential field (illustrating the driving
forces for fluid flow within the plume) for carpets with ratio $\kappa_{\beta}/\kappa_{\alpha}$
greater and less than unity show that the flow is much more complicated
than simple fast/slow channels. This is because a growing plume is
cohesive and incompressible and remains connected to the fluid reservoir.
In no respect does it resemble a collection of diffusing particles.
\begin{acknowledgments}
I thank Professor Robert ``Bob'' Smith (Department of Geological
Sciences) for arranging my access to the resources of the University
of Idaho Library (Moscow, Idaho).
\end{acknowledgments}

\end{document}